\documentstyle[12pt]{article}
\voffset =- 20mm
\newcommand{\be}[1]{\begin{equation}\label{#1}}
\newcommand{\ee}{\end{equation}}
\newcommand{\rf}[1]{(\ref{#1})}
\begin{document}
\title{On Relation between String Theory
       and Multidimensional Cosmology}

\author{Zhuk \\
     Department of Theoretical Physics, University of Odessa,\\
     2 Petra Velikogo Str., 270100 Odessa,
 Ukraine\thanks {email: zhuk@paco.odessa.ua}
}
\maketitle

\abstract{
It is shown that a transition from a multidimensional
cosmological model with one internal space of the dimension
$d_1$ to the effective tree-level bosonic string corresponds to
an infinite number of the internal dimensions:
$d_1\rightarrow\infty$. 
}
%
%
\section{Introduction}

String theories \cite{1} are at the moment the most promising
candidates for a unified description of the basic physical
interactions. The most consistent formulation of these theories
are possible in a space-time with critical dimensions $D_c$ more
than four. For example, $D_c=26$ or $ 10 $ for the bosonic and
supersymmetric version, respectively.

Since string effects become important at Planck scales,
cosmology can provide a natural test for string theories. A lot
of papers were devoted to string cosmology (see e.g. \cite{2} and
references therein).

Another class of models which exploit the idea of extra
dimensions have the metric of the form \cite{3,4}
\be{1.1}
g= -dt\otimes dt e^{2\gamma(t)} +\sum_{i=0}^{n} a_i(t)g^{(i)}
\ee
on the $D$-dimensional manifold
\be{1.2}
M=R\times M_0\times\cdots\times M_n  ,
\ee
where $M_i$ with the metric $g^{(i)}$ are $d_i$-dimensional
spaces of constant curvature (more generally, they are Einstein
spaces). These models are natural generalization of the
Friedmann universe as well as Kasner universe to the
multidimensional case. Classical and quantum dynamics of these
models were considered in many papers (see e.g. \cite{5,6} and
references therein).

It is worth-while to find a relation between string theory (ST)
and multidimensional cosmological models (MCM) of the form
\rf{1.1} - \rf{1.2}.  Here we show that in the case of one internal space
$(n=1)$ the transition from MCM to the effective tree-level
bosonic string occurs in the limit of infinite number of the
internal dimensions:\ $d_1\rightarrow \infty$.

%
\section{String effective action}
\setcounter{equation}{0}

We consider a bosonic string in the presence of a background
consisting of a $D_0$-dimensional metric $g^{(o)}_{\mu \nu}$ and
a dilaton $\Phi$. The action of this string is following
\cite{1} 
\be{2.1}
S_{\sigma} = -\frac{1}{4\pi} \int d^2\sigma \sqrt{h}
\left[
\frac{1}{\alpha^{\prime}} h^{ab} \partial_aX^{\mu} \partial_bX^{\nu}
g^{(o)}_{\mu \nu}(X^{\rho}) 
\right.  
\ee
$$
\left.
+ \Phi(X^{\rho})R^{(2)}
\right]  ,
$$
where $h_{ab}$ is the world-sheet metric tensor, $R^{(2)}$ is
the Ricci scalar constructed with $h_{ab},\ X^{\mu}$ is the
coordinates of the string position and $\alpha^{\prime} $ is the
Regge slope parameter connected with the string tension $T:\
2\pi\alpha^{\prime}T = 1$.

Corresponding tree-level effective action reads \cite{1}
\be{2.2}
S_{eff} = \frac{1}{2\kappa_0^2} \int d^{D_0}x \sqrt{|g^{(0)}|}
e^{-2\Phi} 
\left(
R[g^{(0)}] 
\right.
\ee
$$
\left.
+4 \partial_{\mu}\Phi \partial_{\nu}\Phi g^{(0)\mu\nu}+ C
\right)   ,
$$
where $\kappa_0^2$ is a $D_0$-dimensional  gravitational constant and
$C=-2(D_{eff}-\bar{D})/3\alpha^{\prime}$ is the central charge
deficit  which depends on details of particular ST. For
example, $D_{eff}=D_0,\ \bar{D}=26$ in the bosonic version and
$D_{eff}=\frac{3}{2}D_0,\ \bar{D}=15$ in the supersymmetric
version. The effective action \rf{2.2} is written in the
Brans-Dicke frame. After conformal transformation
\be{2.3}
\hat{g}^{(0)}_{\mu\nu}=e^{-4\Phi/(D_0-2)}g^{(0)}_{\mu\nu},
\ \varphi=\pm\frac{2}{\sqrt{D_0-2}}\Phi 
\ee
the action \rf{2.2} can be rewritten in the Einstein frame as
follows 
\be{2.4}
S_{eff} = \frac{1}{2\kappa_0^2} \int d^{D_0}x \sqrt{|\hat{g}^{(0)}|}
\left(
\hat{R}[\hat{g}^{(0)}] 
\right.
\ee
$$
\left.
-\hat{g}^{(0)\mu\nu} \partial_{\mu}\varphi \partial_{\nu}\varphi
 -2\Lambda e^{-2\lambda_s \varphi} 
\right)   ,
$$
where
\be{2.5}
\lambda_s^2=1/(D_0-2)
\ee
is the string dilatonic coupling constant and
\be{2.6}
\Lambda:=-\frac12 C
\ee
is the " cosmological constant ".

%
\section{Multidimensional cosmological effective action}
\setcounter{equation}{0}

Let us consider now the model \rf{1.1}. We slightly generalize
it to the inhomogeneous case supposing that the metric has the
form 
\be{3.1}
g=g^{(0)} + \sum_{i=1}^{n}a_i^2(x)g^{(i)}   ,
\ee
where the metric $g^{(0)}$ is defined on the
$D_0=d_0+1$-dimensional manifold $\bar{M}_0 = R\times M_0 $ and
$x$ are some coordinates of $\bar{M}_0 $ : $ g^{(0)} =g^{(0)}_{\mu\nu}
dx^{\mu}\otimes dx^{\nu}$.

Hereafter we consider one internal space case $n=1$ with $M_1$
being an Einstein space:
\be{3.2}
R[g^{(1)}]=\lambda^1 d_1.
\ee
The action is taken in the Einstein-Hilbert form
\be{3.3}
S = \frac{1}{2\kappa^2} \int_M d^{D}x \sqrt{|g|}
\left(R[g]-2\Lambda \right) +S_{GM}   ,
\ee
where $S_{GM}$ is the standard Gibbons-Hawking boundary term and
$\kappa^2$ is a $D$-dimensional gravitational constant. The
cosmological effective action is obtained by the dimensional
reduction of the action \rf{3.3} and reads \cite{7}
\be{3.4}
S_{c} = \frac{1}{2\kappa_0^2} \int_{\bar{M}_0} d^{D_0}x \sqrt{|g^{(0)}|}
e^{-2\Phi} 
\left(
R[g^{(0)}] 
\right.
\ee
$$
\left.
-4 \omega \partial_{\mu}\Phi \partial_{\nu}\Phi g^{(0)\mu\nu}
+ R[g^{(1)}]e^{\frac{4}{d_1}\Phi}-2\Lambda
\right)   ,
$$
where the dilaton $\Phi$ is defined via the scale factor
$a_1(x)$ as follows
\be{3.5}
e^{-2\Phi}:=a_1^{d_1}.
\ee
The action \rf{3.4} is written in the Brans-Dicke frame with
the Brans-Dicke parameter
\be{3.6}
\omega=\frac{1}{d_1}-1\ <\ 0.
\ee
It follows from \rf{3.4} that this action in the limit
$d_1\rightarrow\infty$ turns into the string effective action
\rf{2.2} with $C:=R[g^{(1)}]-2\Lambda$.

If $D_0=2$ the equation \rf{3.4} represents $2D$ dilaton gravity
obtained from inhomogeneous cosmology.
There is no Einstein frame for $2D$ manifolds. This is not a
fault of the theory but rather corresponds to the wellknown fact
that 2-dimensional Einstein equations are empty, i.e. they do
not imply a dynamics. We can see it explicitly from the
conformal transformations \rf{2.3} which are singular for
$D_0=2$. But for $D_0\neq 2$ we can obtain the cosmological
action \rf{3.4} in the Einstein frame.

By analogy, with the conformal transformations \rf{2.3} we can
write 
\be{3.7}
\hat{g}^{(0)}_{\mu\nu}=e^{-4\Phi/(D_0-2)}g^{(0)}_{\mu\nu},
\ee
$$
\varphi=\pm 2\left[ \omega+ \frac{D_0-1}{D_0-2} \right]^{1/2}\Phi.
$$
The action \rf{3.4} then reads
\be{3.8}
S_{c} = \frac{1}{2\kappa_0^2} \int_{\bar{M}_0} d^{D_0}x 
\sqrt{|\hat{g}^{(0)}|}
\left\{
\hat{R}[g^{(0)}]
\right.
\ee
$$
\left.
 -\hat{g}^{(0)\mu\nu} \partial_{\mu}\varphi
\partial_{\nu}\varphi +R[g^{(1)}]e^{-2\lambda_{(1)c}\varphi} -2\Lambda
e^{-2\lambda_{(2)c}\varphi}
\right\}   ,
$$
where the cosmological dilatonic coupling constants are
\be{3.9}
\lambda^2_{(1)c}= \frac{D-2}{d_1(D_0-2)},
\lambda^2_{(2)c}= \frac{d_1}{(D-2)(D_0-2)},
\ee
here $D=D_0+d_1=1+d_0+d_1$ and both of these coupling constants
go to the string dilatonic coupling constant $\lambda_s$
\rf{2.5} in the limit $d_1\rightarrow\infty$.

In conclusion we would like to note that the case of more than
one internal space $(n >1)$ corresponds to multidilatonic
theories. This type of the multidimensional cosmological models
may be related to the very fashionable now $M$-theory \cite{8}
but this question needs more detailed investigations.

Here, we considered the multidimensional cosmology regardless of
strings and after dimensional reduction only the connection between them
was found. However, MCM may have its origin directly from ST. 
Actually, the action \rf{3.3} may be considered as the tree-level
effective action \rf{2.4} for the non-critical bosonic string with
a constant background dilaton: $\Phi = const$ . For the critical
string $(D_0 = D_c)$ with $\Phi \ne const$ we obtain the Einstein
action with free scalar field. If we propose now that metric in the
Einstein frame has the form \rf{1.1} then we obtain MCM. This form of 
the  metric is very convenient for investigation of the dynamical
reduction problem in ST. Of special interest are exact solutions
because they can be used for a detailed study of the compactification
of the internal spaces \cite{6} .

\vspace{5mm}
\noindent Acknowledgement\\
\noindent The author is grateful to V.D.Ivashchuk for useful
discussions. 

%

%
\end{document}